\documentclass[aps,prx,twocolumn,superscriptaddress,noshowpacs]{revtex4-2}
\usepackage{graphicx}
\usepackage{color}
\usepackage{float}
\usepackage{bm}
\usepackage{braket}
\usepackage{amsmath}
\usepackage{lipsum}

\usepackage[normalem]{ulem}

\newcommand{\bk}{{\bf k}}

\newcommand{\bq}{{\bf q}}

\newcommand{\eye}{\mathrm{i}}

\newcommand{\comment}[1]{}

\begin{document}

\title{Quantum fluctuations of charge order induce phonon softening in a superconducting cuprate}


\author{H. Y. Huang}
\affiliation{National Synchrotron Radiation Research Center, Hsinchu 30076, Taiwan}

\author{A. Singh}
\affiliation{National Synchrotron Radiation Research Center, Hsinchu 30076, Taiwan}

\author{C. Y. Mou}
\affiliation{Center for Quantum Technology and Department of Physics, National Tsing Hua University, Hsinchu 30013, Taiwan}

\author{S. Johnston}
\affiliation{Department of Physics and Astronomy, The University of Tennessee, Knoxville, TN 37996, USA}

\author{A. F. Kemper}
\affiliation{Department of Physics, North Carolina State University, Raleigh, NC 27695, USA}

\author{J. van den Brink}
\affiliation{Institute for Theoretical Solid State Physics, IFW Dresden, Helmholtzstrasse 20, D-01069 Dresden, Germany}

\author{P. J. Chen}
\affiliation{Department of Mechanical Engineering, City University of Hong Kong, Kowloon, Hong Kong}
\affiliation{Hong Kong Institute for Advanced Study, City University of Hong Kong, Kowloon, Hong Kong}

\author{T. K. Lee}
\affiliation{Institute of Physics, Academia Sinica, Taipei 11529, Taiwan}
\affiliation{Department of Physics, National Sun Yat-sen University, Kaohsiung, 80424, Taiwan}

\author{J. Okamoto}
\affiliation{National Synchrotron Radiation Research Center, Hsinchu 30076, Taiwan}
\author{Y. Y. Chu}
\affiliation{National Synchrotron Radiation Research Center, Hsinchu 30076, Taiwan}

\author{J. H. Li}
\affiliation{Department of Physics, National Tsing Hua University, Hsinchu 30013, Taiwan}
\affiliation{National Synchrotron Radiation Research Center, Hsinchu 30076, Taiwan}

\author{S. Komiya}
\affiliation{Central Research Institute of Electric Power Industry, Yokosuka, Kanagawa, 240-0196, Japan}

\author{A. C. Komarek}
\affiliation{Max Planck Institute for Chemical Physics of Solids, Nöthnitzerstrasse 40, 01187 Dresden, Germany}

\author{A. Fujimori}
\affiliation{Department of Applied Physics, Waseda University, Shinjuku-ku, Tokyo 169-8555, Japan.}
\affiliation{National Synchrotron Radiation Research Center, Hsinchu 30076, Taiwan}

\author{C. T. Chen}
\affiliation{National Synchrotron Radiation Research Center, Hsinchu 30076, Taiwan}

\author{D. J. Huang}
\altaffiliation [email: ] {\emph{djhuang@nsrrc.org.tw}} 
\affiliation{National Synchrotron Radiation Research Center, Hsinchu 30076, Taiwan}
\affiliation{Department of Physics, National Tsing Hua University, Hsinchu 30013, Taiwan}


\begin{abstract}
Quantum phase transitions play an important role in shaping the phase diagram of high-temperature cuprate superconductors. These cuprates possess intertwined orders which interact strongly with superconductivity. However, the evidence for the quantum critical point  associated with the charge order in the superconducting phase remains elusive. Here we show the short-range charge orders and the spectral signature of the quantum fluctuations in La$_{2-x}$Sr$_x$CuO$_4$ (LSCO) near the optimal doping using high-resolution resonant inelastic X-ray scattering. On performing calculations through a  diagrammatic framework, we discovered that the charge correlations significantly soften several branches of phonons. These results elucidate the role of charge order in the LSCO compound, providing evidence for quantum critical scaling  and discommensurations associated with charge order.
\end{abstract}

\date{\today}

\maketitle

\thispagestyle{empty}

\section{Introduction}

When doped with holes or electrons,  cuprates at low temperatures can be tuned from a Mott insulating phase to a superconducting phase, and then to a Fermi liquid phase. In addition, an enigmatic  pseudogap phase exists in the underdoped regime of hole-doped cuprates above the superconductivity transition temperature $T_{_{\rm C}}$ with a crossover temperature $T^*$, which decreases monotonically when the doping is increased \cite{KeimerNature2015}. Several symmetry-breaking orders such as charge-density waves (CDW) have also been discovered in 
the cuprates with comparable onset temperatures \cite{FradkinRMP2015}.

Charge fluctuations are one of the most fundamental collective excitations in matter.  Recently, CDW in cuprate superconductors have attracted renewed interest \cite{GhiringhelliScience2012,Blackburn2013,Croft2014, zhu2015,CominARCMP2016,MiaoPRX2019,WenNatComm2019,ArpaiaScience2019,Mitrano2019,Torchinsky2013,CominScience2014,Tabis2014,daSilvaNeto2015,Campi2015,Gerber2015,Peng2018,Kim2018,Chen2019,Yu2020,ChaixNatPhys2017,LeeNatPhys2020,LinPRL2020,LiPNAS2020}
but their mechanism and competition with superconductivity remain subjects of vigorous discussion. 
The wave vector of the CDW, for example, exhibits an inconsistent evolution with doping in various cuprate families \cite{CominARCMP2016,MiaoPRX2019}. This observation has lead to debate on whether real-space local interactions or Fermi surface nesting in momentum space underlies the CDW physics.  Moreover, CDW are coupled to phonons in cuprates \cite{Castellani1995,Becca1966,fukuda2005,ReznikNature2006,giustino2008,LeTacon2014,ChaixNatPhys2017,LeeNatPhys2020,LinPRL2020,ParkPRB2014,LiPNAS2020,Miao2021}, but the underlying mechanism of phonon softening remains a mystery \cite{zhu2015}, although lattice vibrations in metals can be damped or softened by electronic quasiparticles or excitations.

Upon tuning a non-thermal parameter through a critical value, quantum phase transitions occur at the absolute zero of temperature $T$. The  putative quantum critical point (QCP) in a cuprate holds the key to understanding many profound phenomena related to its superconductivity \cite{sachdev2010,sachdev2011,CapraraPRB2017,proust2019,Michon2019,Cooper2009,LegrosNatPhys2019}. 
Anomalous thermodynamic \cite{Michon2019} and transport properties \cite{Cooper2009,LegrosNatPhys2019} of cuprates close to a QCP have been observed. For example, as temperature $T$ approaches zero, the electrical resistivity varies linearly with $T$, rather than the $T^2$ dependence of a Fermi liquid.
In approaching a QCP, two quantum states of the system can exchange their orderings in energy. As a result, short-range CDW states can be intertwined with the superconducting state. Also, because the energy difference is small, switching back and forth between CDW and superconducting ground states is energetically likely, resulting in quantum charge fluctuations.
Interestingly, recent resonant inelastic X-ray scattering (RIXS) studies have found that the charge-density fluctuations permeate through a broad region of the cuprate phase diagram \cite{ArpaiaScience2019}. 

A  crucial task in the scenario of the presence of QCP is to identify the order phase that terminates at the QCP. In the present work, we shall presume the order phase to be the CDW order and try to find evidence for the QCP associated with the CDW in the superconducting phase. 
Inelastic X-ray scattering (IXS) probes charge fluctuations by measuring the dynamical structure factor $S({\bf q}, \omega)$, which is the space and time Fourier transformation of the density-density correlation function. Here, ${\hbar\bf q}$ and ${\hbar\omega}$ are the momentum and energy transferred to charge excitations, respectively, with $\hbar$ being the Planck constant $h$ divided by $2\pi$. For a given electron system of charge density $n ({\bf r}, t)$ at position ${\bf r}$ and time $t$, the charge fluctuation is $\delta n({\bf r}, t)= n({\bf r}, t) -  \langle n({\bf r}, t)\rangle$, in which $\langle \cdots\rangle$ denotes the quantum statistical average. The density-density correlation function $\langle  n({\bf r}, t)  n({\bf r}', t')\rangle$ can then be written as 
 \begin{equation}
\langle n({\bf r}, t) n({\bf r}^\prime, t^\prime)\rangle=\langle n({\bf r}, t)\rangle \langle n({\bf r'}, t')\rangle + \langle \delta n({\bf r}, t) \delta n({\bf r}^\prime, t^\prime)\rangle.
\end{equation}
In the presence of disorder, an additional average over disorders must be taken over Eq.~(1).  X-ray scattering data thus comprise contributions of the static charge distribution $\langle n({\bf r}, t)\rangle \langle n({\bf r'}, t')\rangle$ and its dynamical fluctuations $\langle \delta n({\bf r}, t) \delta n({\bf r}^\prime, t^\prime)\rangle$, corresponding to charge susceptibilities denoted by $\chi_0 ({\bf q}, {\omega})$ and $\chi({\bf q}, {\omega})$, respectively. 
In addition, through the fluctuation-dissipation theorem, $S({\bf q}, \omega)$ is related to the charge susceptibility $\chi({\bf q}, {\omega})$ by $S({\bf q},\omega) =2{\hbar}(1-e^{-\beta\hbar\omega})^{-1} {\rm Im} \chi({\bf q}, {\omega})$, in which $\beta = 1/k_{_B}T$, with $T$ and $k_{_B}$ denoting temperature and the Boltzmann constant, respectively. 

O $K$-edge RIXS probes certain characteristics of the dynamical charge fluctuations of CDW, although additionally modulated by the effects of RIXS matrix elements, light polarization, and orbital characters \cite{Ament11,ArpaiaScience2019,LeeNatPhys2020,JiaPRX2016}. 
In this work, we have performed high-resolution O $K$-edge RIXS measurements and theoretical calculations through a  diagrammatic framework to investigate the quantum fluctuations of charge order in superconducting cuprate La$_{2-x}$Sr$_x$CuO$_4$ (LSCO) near the optimal doping. Our results elucidate the role of  charge order in the LSCO compound, providing evidence for  the quantum critical scaling associated with charge order.


\begin{figure}[t]
\centering
\includegraphics[width=8.6 cm]{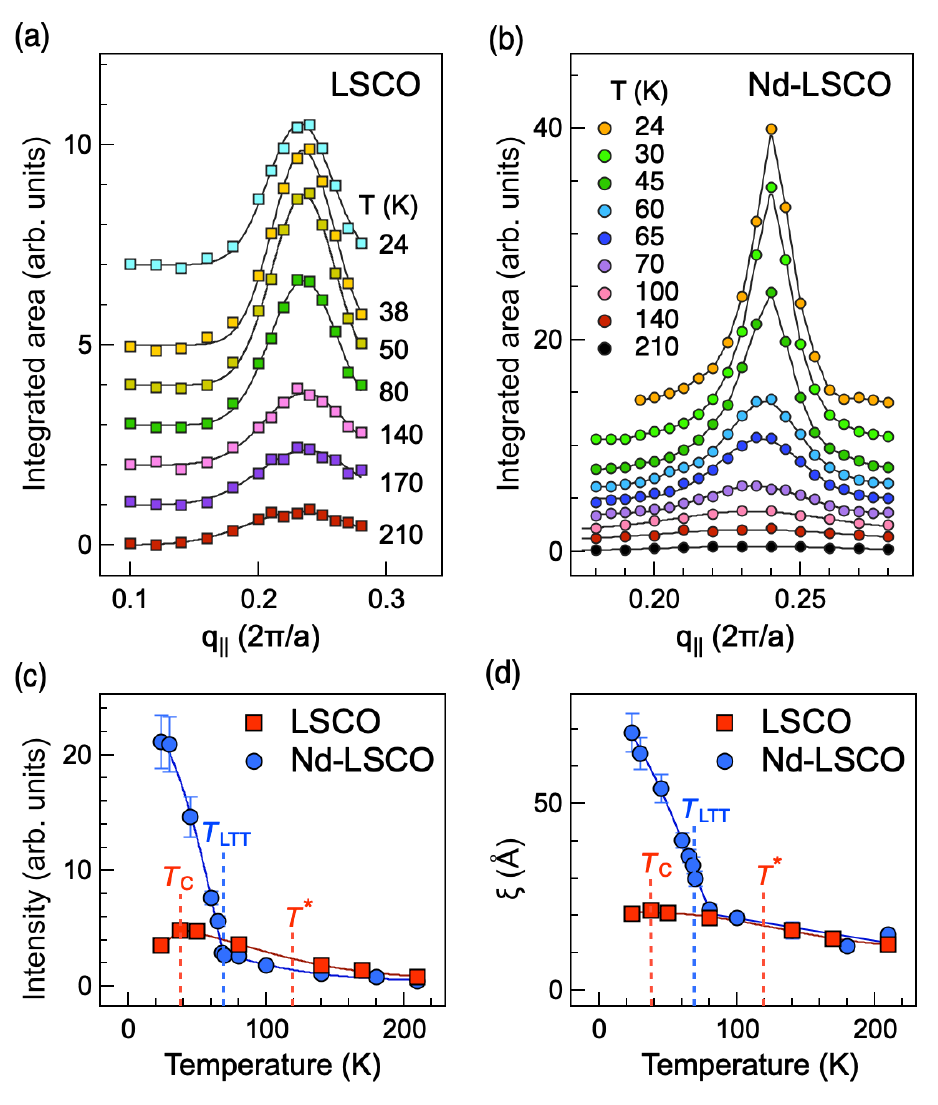}
\caption{ O $K$-edge elastic scattering of La-214 cuprates. {(a)}~\&~{(b)}, scattering intensities of LSCO  and Nd-LSCO vs. in-plane momentum \textbf{q}$_{\|}$ along the antinodal direction $(\pi, 0)$ at various temperatures.  Energy-resolved elastic-scattering plots were extracted from the integrated area of O $K$-edge RIXS over the energy from $-5$ meV to $5$ meV. The incident X-ray energy for the RIXS spectra was tuned to the mobile hole of the so-called Zhang-Rice singlet (ZRS) with an absorption energy near 528.5 eV. All spectra are vertically offset for clarity. 
{(c)}~\&~{(d)}, CDW intensity and correlation length ($\xi$) of LSCO and Nd-LSCO as a function of temperature, respectively.  The vertical dashed lines show $T_{_{\rm C}}$ and $T^*$ of LSCO, and structural transition temperature $T_{\rm LTT}$ of Nd-LSCO. All solid lines serve as visual guidance.}
\end{figure}

\section{RIXS measurements} 

\subsection{Experimental} 

The LSCO single crystals with the doping level {$x=0.15$} were grown by the traveling-solvent floating zone method \cite{Komiya2002, Komiya2005, Ono2007}. After growth, the crystals were annealed to remove oxygen defects. The value of $x$ was determined from an inductively-coupled-plasma atomic-emission spectrometric analysis. The $T_{_{\rm C}}$ of the $x=0.15$ sample is 37.5~K.

We conducted O $K$-edge RIXS measurements using the AGM-AGS spectrometer of beamline 41A at Taiwan Photon Source of National Synchrotron Radiation Research Center, Taiwan \cite{Singh2021}. This recently constructed AGM-AGS beamline is based on the energy compensation principle of grating dispersion. High-resolution RIXS data were measured with an energy resolution of 16 meV at incident photon energy of 530~eV. See Supplementary Material for the details of the AGM-AGS spectrometer and data analysis.

\begin{figure*}[t!]
\centering
\includegraphics[width=2\columnwidth]{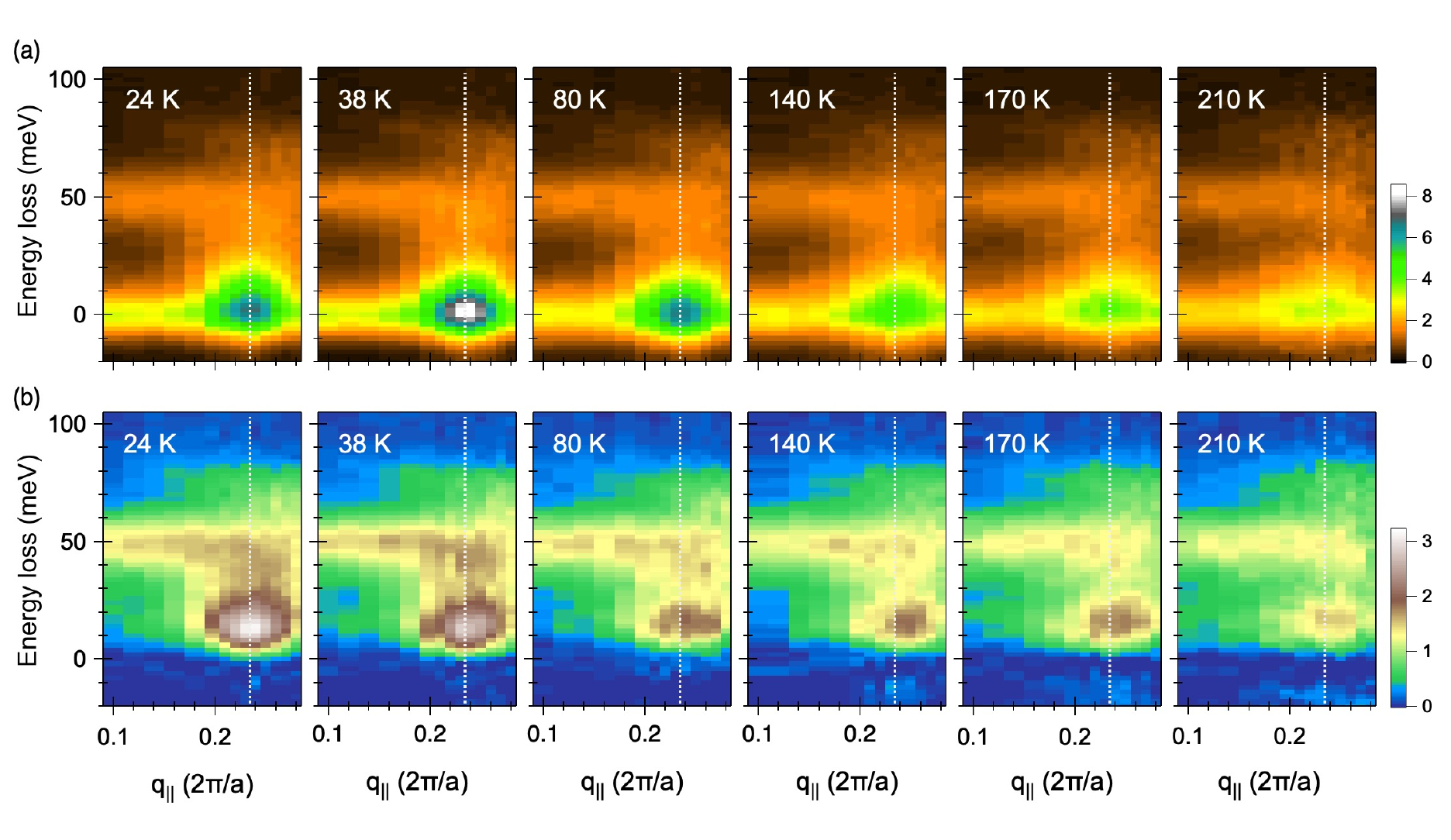}
\caption{Temperature-dependent O $K$-edge RIXS of optimally doped LSCO. {( a)}, RIXS intensity distribution maps in the plane of energy loss vs. in-plane momentum transfer \textbf{q}$_\|$ along $(\pi, 0)$ at various temperatures. {(b)}, RIXS intensity distribution maps after the subtraction of elastic scattering. All RIXS spectra were recorded with $\sigma$-polarized incident X-rays of energy tuned to the ZRS hole. The momentum transfer is \textbf{q}~$=(q_{\|}, 0, L)$ with $L$ varying between 0.47 and 1.03 in reciprocal lattice units.  Dotted lines indicate the position of measured \textbf{q}$_{\rm CDW}~=(0.23, 0, L)$.}
\end{figure*}

\subsection{Charge order of LSCO} 

We first measured static charge distributions $\langle n({\bf r}, t)\rangle \langle n({\bf r^\prime}, t^\prime)\rangle$ of LSCO with ${x~=~0.15}$ using energy-resolved elastic X-ray scattering. Figure 1(a) plots the intensity of O $K$-edge elastic scattering as a function of in-plane wave vector change \textbf{q}$_{\|}$ varied along the anti-nodal direction $(\pi, 0)$ at various temperatures.  We observed CDW correlations in LSCO with an in-plane modulation vector ${\bf q}_{_{\rm CDW}} = (0. 235, 0)$, which is given in reciprocal lattice units (r.l.u.) throughout this paper. Figures 1(c) and 1(d) show, respectively, the temperature-dependent CDW intensity and correlation length ($\xi$) defined as the inverse of the half width at half maximum (HWHM) of the momentum scan. The charge order of optimally doped LSCO exhibits a short correlation length, ranging from 22~{\AA} to 12~{\AA}. Consistent with previous results of LSCO, the CDW are slightly suppressed in the superconducting phase as the temperature is decreased across $T_{_{\rm C}}$ \cite{Croft2014,WenNatComm2019}.

The observed charge correlations of LSCO persist over a wide temperature range, up to well above  $T^*$ and into the strange-metal phase. These correlations 
achieve no long-range charge order, however, in contrast with other La-214 cuprates. For example, La$_{1.48}$Nd$_{0.4}$Sr$_{0.12}$CuO$_{4}$ (Nd-LSCO), which has a low-temperature tetragonal (LTT) structure, shows long-range and short-range charge orders, as plotted in Fig 1(b).  
The short-range charge order of LSCO might be caused by spatial disorders or phase separation arising from a subtle balance between Coulomb interactions and kinetic energy. At this time we cannot determine if the CDW with ${\bf q}_{_{\rm CDW}} = (0. 235, 0) $ are truly incommensurate or instead appear incommensurate due to defects known as discommensurations in the commensurate CDW state \cite{McMillanPRB1976,ChenPRL1981,TuSciRep2019,MesarosPNAS2016}. In a discommensurated structure, locally commensurate CDW regions are separated by discommensurations within which the CDW phase varies rapidly. 

To unveil the charge instability and the coupling between CDW fluctuations and phonons in LSCO, we performed high-resolution O $K$-edge RIXS measurements. Figure~2(a) presents a series of temperature-dependent RIXS intensity distribution maps in the plane of energy loss vs. in-plane momentum transfer \textbf{q}$_\|$ along $(\pi, 0)$; Fig.~2(b) shows those maps after the subtraction of elastic scattering.
There exists a pronounced excitation feature of energy about 14~meV in the neighborhood of ${\bf q}_{_{\rm CDW}}$. Its intensity decreases with increasing temperature, unlike that of the static CDW intensity that has a maximum at $T_{_C}$ as plotted in Fig.~1(c) with red squares. This low-energy RIXS excitation appears to shift away from $q_{_{\rm CDW}}$ when the thermal energy approaches the excitation energy. This reflects that, at low temperatures, the low-energy RIXS excitation is dominated by charge-density fluctuations and its contribution from the coupling of CDW to acoustic phonons increases with the increase of temperature, as discussed later.

\begin{figure}[t]
\centering
\includegraphics[width=8.0 cm]{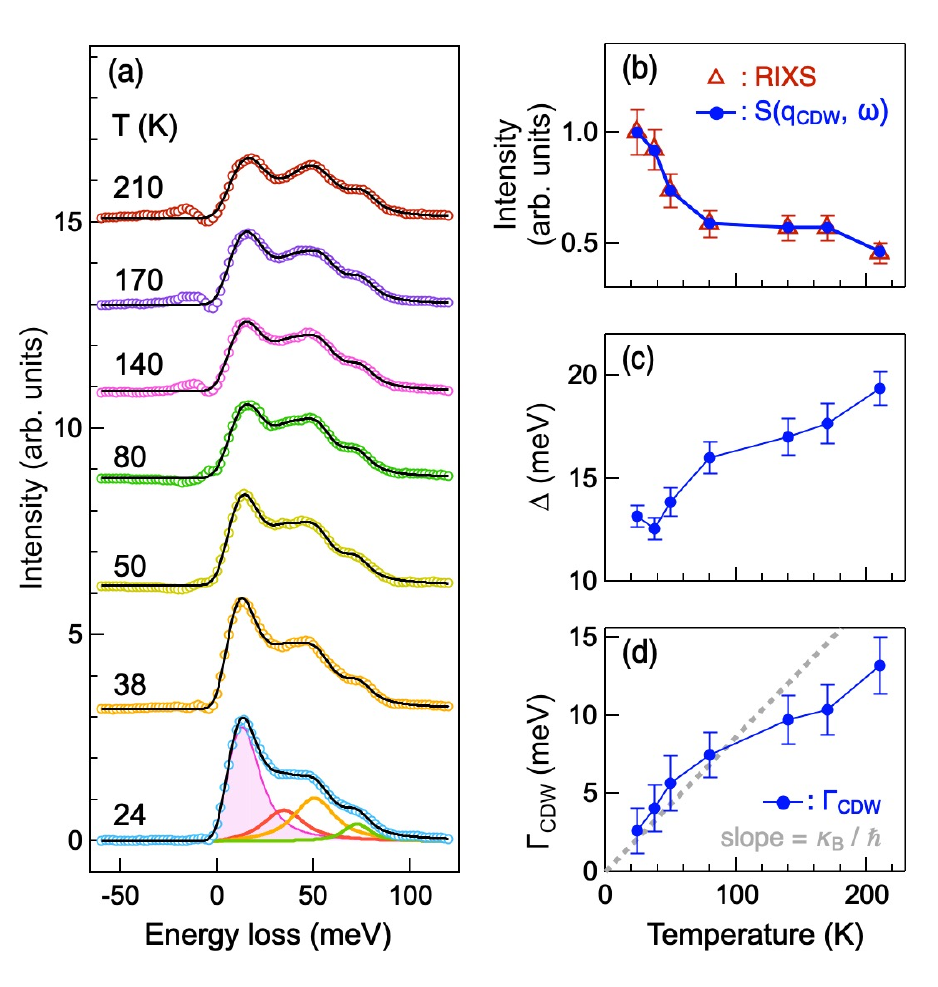}
\caption{Temperature dependence of CDW fluctuations. (a), 
Momentum-integrated RIXS spectra after the subtraction of elastic scattering and then momentum integration over a range from 0.22 to 0.25 in units of ${2\pi}/a$ at different temperatures. The RIXS spectra were fitted to four phonon modes. The component derived from the coupling of CDW fluctuations to acoustic phonons is shaded in pink.  The components of the BS, $A_{1g}$ and $B_{1g}$ phonon modes are shown in green, orange and red, respectively.  Details of the curve fitting are presented in Supplementary Material. 
{(b)}, RIXS {peak intensity  of the 14-meV component shown in (a)}, compared to the maximum of calculated $S({\bf q}_{_{\rm CDW}}, \omega)$ { after convolution with the instrumental resolution.} The value of $S({\bf q}_{_{\rm CDW}}, \omega)$ with the 24-K data point is normalized to the RIXS intensity.
Values of $\Delta$ and $\Gamma_{_{CDW}}$ used for the $S({\bf q}_{_{CDW}}, \omega)$ calculations are plotted in {(c)} and {(d)}, respectively. {(d)}, Temperature dependence of $\Gamma_{_{CDW}}$.  The gray dashed line indicates the slope of the lifetime width vs. $T$  determined by the limit of the Planckian dissipation. All solid lines connecting data points in (b) - (d) serve as visual guidance. }
\end{figure} 

\subsection{CDW fluctuations}

For a system which exhibits a QCP associated with charge order,  $S({\bf q}, \omega)$ near ${\bf q}_{_{\rm CDW}}$ measures the collective fluctuations sensitive to the proximity of QCP. When the thermal energy $k_{_B}T$ is less than the typical energy of order parameter fluctuations and less than energy $\hbar\omega$, their quantum nature is important. For low temperatures, we expect that the finite frequency fluctuation is dominated by the amplitude fluctuation of the charge density, and the following phenomenological form for charge susceptibility describes the quantum fluctuations \cite{sachdev2011,ArpaiaScience2019}:
\begin{equation} \label{chi}
\chi_{_{\rm CDW}}({\bf q}_{\|}, {\omega})=  \frac{1}{\Delta^2+c^{2}({\bf q}_{\|}-{\bf q}_{_{\rm CDW}})^2-(\omega +i\Gamma_{_{\rm CDW}})^2}.
\end{equation}
Here, $\Delta$ is the characteristic energy of the fluctuations at $T=0$ and can be interpreted as the inverse
correlation length for the amplitude fluctuations. One expects that the characteristic energy $\Delta$ vanishes when the doping level approaches the quantum critical point. $c$ is a parameter that characterizes the speed of the excitations with dispersion $\omega \sim  c |{\bf q}_{\|} -{\bf q}_{_{\textrm{CDW}}}|$, and $\Gamma_{_{\rm CDW}}$ is a parameter that characterizes the lifetime of the CDW amplitude excitations.  In contrast, the static charge susceptibility $\chi_0 ({\bf q}, {\omega})$ vanishes except for $\omega \sim 0$ and its nearby region of CDW phase fluctuations characterized by Eq.(\ref{chi}) with $\Delta=0$ and parameters $c$ and $\Gamma_{_{\rm CDW}}$. This energy region is beyond our RIXS energy resolution.

Figure 3(a) plots RIXS spectra after the subtraction of elastic scattering for momentum integrated from $q_{\|}=0.22$ to $q_{\|}=0.25$ at different temperatures. This low-energy RIXS intensity decreases when the temperature is increased as plotted in Fig. 3(b). On comparing the intensity, spectral profile, and peak energy of the low-energy RIXS excitation near 14~meV with those of calculated $S({\bf q}_{_{\rm CDW}}, {\omega})$,  in which the contribution of the dispersion parameter $c$ vanishes, one can obtain the evolution of $\Delta$ and $\Gamma_{_{\rm CDW}}$ for various temperatures. Figures 3(c) and 3(d) show the estimated temperature-dependent $\Delta$  and $\Gamma_{_{\rm CDW}}$,  respectively. The evolution of $\Delta$, which is inversely proportional to  the CDW correlation length \cite{CapraraPRB2017}, is accordant with the temperature dependence of static charge order shown in Fig. 1. 
At low temperatures, $\Gamma_{_{\rm CDW}}$ grows with increasing temperature. In the quantum critical regime,  the scattering rate of quasi-particles reaches the Planckian limit and  the change of $\Gamma_{_{\rm CDW}}$ is comparable with  ${k_{_B}T}/\hbar$, i.e., the Planckian dissipation \cite{zaanen2004,LegrosNatPhys2019}. Although the measured data cover only one temperature point below $T_{_{\rm C}}$, this observation supports the quantum critical nature of CDW fluctuations in LSCO.  

Figure 3(d) also shows that, as $T$ is increased, the temperature dependence of $\Gamma_{_{\rm CDW}}$ exhibits a bending feature about $80~K$. This suggests two  possible  scenarios: a crossover near the QCP and the coupling of CDW fluctuations to acoustic phonons. First, a crossover of $\Gamma_{_{\rm CDW}}$ behavior near the QCP \cite{chakravarty1989, chakravarty2019} indicates that Planckian linear-T dependence works only at the critical regime (low temperature regions) controlled by the QCP. In other words, only inside the critical regime,  $\Gamma_{_{\rm CDW}}$  shows the linear-T dependence, which reflects the gapless nature at the QCP, i.e., $1/\Gamma_{_{\rm CDW}}$  going to infinity. At higher temperatures beyond the critical regime, $\Gamma_{_{\rm CDW}}$ no longer reaches the Planckian limit. In this scenario,  the quantum critical region is up to $\sim~80~K$.
In the second scenario, the bending feature of $\Gamma_{_{\rm CDW}}$ suggests that, for the thermal energy $k_{_B}T$ comparable to the charge fluctuation energy, CDW fluctuations alone do not explain the RIXS data at high temperatures, as shown in supplementary Figs. S8(e) - S8(g). This indicates that the low-energy RIXS is also contributed by the coupling of CDW fluctuations to phonons. We hence calculated the full RIXS intensity using a diagrammatic framework, in which the coupling between $\chi_{_{\rm CDW}}$  and phonons renormalizes the phonon propagator.


\section{RIXS calculations}

The intensity of the phonon excitations was calculated by using the diagrammatic 
framework \cite{Devereaux2016} generalized to the O $K$-edge. 
The RIXS intensity is given by ($\hbar = 1$)

\begin{widetext}

\begin{equation}
\begin{aligned} 
    I_\nu({\bf q}, \Omega)&=\frac{1}{N}\sum_{\bk,\alpha}\frac{1}{N}\sum_{\bk^\prime,\beta} 
    [\hat{\boldsymbol{\epsilon}}_i \cdot \hat{\mathbf{r}}_\alpha]
    [\hat{\boldsymbol{\epsilon}}_f \cdot \hat{\mathbf{r}}_\beta]\\ 
    &\times g_\nu(\bk,\bk+\bq)
    |\phi_{2p_\alpha}(\bk)|^2|\phi_{2p_\beta}(\bk+\bq)|^2 
    \left[\frac{n_f(-\epsilon_\bk)}{\epsilon_\bk-\omega_f - E_{1s,2p}-\eye \Gamma}-\frac{n_f(-\epsilon_{\bk+\bq})}{\epsilon_{\bk+\bq}-\omega_i - E_{1s,2p}-\eye \Gamma}\right]\\
    &\times
    g_\nu(\bk^\prime+\bq,\bk^\prime)
    |\phi_{2p_\beta}(\bk^\prime)|^2|\phi_{2p_\alpha}(\bk^\prime+\bq)|^2 
    \left[\frac{n_f(-\epsilon_{\bk^\prime})}{\epsilon_{\bk^\prime}-\omega_f - E_{1s,2p}+\eye \Gamma}-\frac{n_f(-\epsilon_{\bk^\prime+\bq})}{\epsilon_{\bk^\prime+\bq}-\omega_i - E_{1s,2p}+\eye \Gamma}\right]\\ 
    &\times
    -\frac{1}{\pi}\mathrm{Im}\left\{\frac{1}{\epsilon_\bk-\epsilon_{\bk+\bq}+\Omega+\eye\gamma_e }D_\nu(\bq,\Omega) \frac{1}{\epsilon_{\bk^\prime}-\epsilon_{\bk^\prime+\bq}+\Omega-\eye \gamma_e}\right\}.
    \label{Eq:ModeIntensity}
\end{aligned}
\end{equation}
\end{widetext}
Here, $n_f(x)$ is the Fermi factor; $\epsilon_\bk$ is the electron band dispersion; $g_\nu(\bk,\bq)$ and $D_\nu(\bq,\xi)$ 
are the electron-phonon vertex and phonon propagators for phonon branch $\nu$, respectively; $\phi_{2p_\alpha}(\bk)$ is the oxygen $2p_\alpha$ orbital character of the band; $\hat{{\bf r}}_\alpha$ is a unit vector pointing along the direction of the 
oxygen orbital; $\Gamma$ is the core-hole lifetime parameter; $\gamma_e$ is a lifetime parameter for the electrons; and $\Omega = \omega_i-\omega_f$ and $\bq = \bk_i-\bk_f$ are the energy and momentum transferred to the sample. 
The phonon softening is determined by the total electronic polarizability. Our model essentially assumes that the phonons respond to some strong CDW fluctuation caused by an external source like correlations. That is, the coupling of CDW fluctuations to phonons renormalizes the phonon propagator through charge susceptibility  $\chi_{_{\rm CDW}}({\bf q}_{\|}, {\omega})$. See Supplementary Material for the calculation details.

Figure 4(a) presents the calculated RIXS intensity distribution map. That the coupling of CDW  fluctuations to acoustic phonons enhances the RIXS intensity significantly at 14~meV for momentum about ${\bf q}_{_{\rm CDW}}$ agrees satisfactorily well with the measurements shown in Fig.~4(b), corroborating the  phenomenological form of charge susceptibility given in Eq.~(2). 

\begin{figure}[ht]
\centering
\includegraphics[width= 8.6 cm]{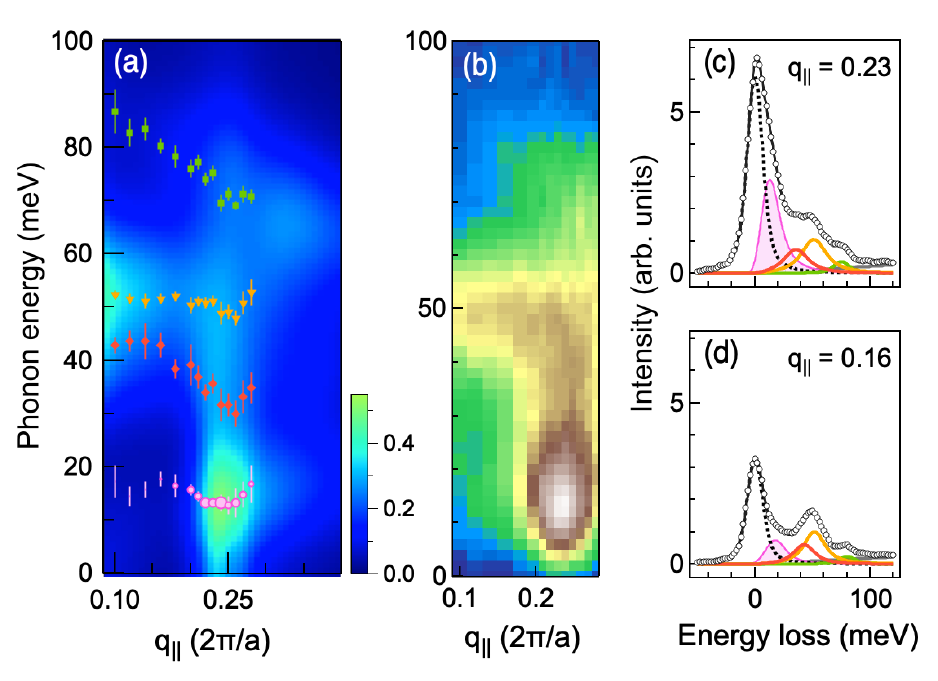}
\caption{{Phonon excitations from RIXS measurements of optimally doped LSCO}.  {(a)}, Calculated intensity distribution map using the diagrammatic 
framework in comparison with phonon dispersions obtained from RIXS spectra. Green squares, orange triangles and red diamonds depict the dispersions of BS, $A_{1g}$ and $B_{1g}$ phonon modes, respectively. The dispersions of the low energy RIXS excitations are represented by pink circles scaled to the fitted intensity. {(b)}, RIXS intensity distribution map of $T = 24$~K extracted from Fig. 2(b) and plotted with the same color scale.  {(c)}$ \& ${(d)}, Measured RIXS spectra with fitted components for phonon excitations of momenta $q_{\|}=0.23$ and $0.16$ at $T = 24$~K, respectively. The curve fitting scheme is same as that of Fig. 3(a). RIXS data are plotted as open circles with black curves to show the summation of fitted components and a linear background. The component derived from the coupling of CDW fluctuations to acoustic and optical phonons is shaded in pink.  The components of the BS, $A_{1g}$ and $B_{1g}$ phonon modes are shown in green, orange and red, respectively. The elastic component is plotted as a dashed curve.}
\end{figure} 

\section{electron-phonon coupling}

In addition to the pronounced low-energy feature, 
RIXS also measures other phonon excitations \cite{Devereaux2016,Ament11a}. The electron-phonon coupling involved with states near the Fermi level is weak for the apical mode \cite{Johnston2010}. The contribution of apical phonons to RIXS can be neglected. We therefore fitted RIXS spectra with three components for high-energy phonons, i.e., bond-stretching (BS), $A_{1g}$ buckling and $B_{1g}$ buckling phonons \cite{mcqueeney1999,giustino2008,sugai2013,ParkPRB2014,Zhou2005}, as shown in Fig.~4(c) and 4(d), which plot fitting spectra of  $q_{\|} = 0.23$ and 0.16 at 24~K, respectively. Results of other $q_{\|}$ and temperatures are presented in Figs.~S9,~S10 and~S11 in Supplementary Material. Figure 4(a) also depicts the phonon dispersions from curve fitting. The diagrammatic calculations explain well the observed photon softening. In addition to phonon softening, the phonon energies from RIXS agree with those are from density-functional-theory (DFT) calculations \cite{giustino2008} and measurements of Raman scattering \cite{sugai2013}, inelastic neutron scattering \cite{mcqueeney1999}, IXS \cite{fukuda2005} and photoemission \cite{Zhou2005}.  We also fitted the RIXS data of $T=24$~K with three phonon modes, i.e., BS, buckling and acoustic phonons; the fitting results plotted in supplementary Fig.~S12 show a phonon softening consistent with that from the four-phonon fitting.  That is, our curve fitting results show that these phonons all exhibit the strongest softening at ${\bf q}_{\|}=(0.25, 0)$ rather than  ${\bf q}_{_{\rm CDW}}$, supporting the scenario of discommensurate CDW in LSCO at doping level $x = 0.15$. In domains of discommensurate CDW,  phonons are strongly coupled to charge fluctuations and the phase shift between scattered X-rays from separate domains is negligible. The phonon softening therefore occurs at ${\bf q}_{\|}=(0.25, 0)$ for CDW domains of period $4a$, in which $a$ is the in-plane lattice constant. 

To further understand the characteristics of CDW in LSCO, we examined the  coupling of CDW to phonons through performing DFT calculations in the scheme of frozen phonons, which were adopted to study qualitatively the coupling of longitudinal BS mode and apical phonon mode of $A_{1g}$ symmetry with a periodicity of $4a$. See Supplementary Material for the details. 
The calculations show that the apical mode has an energy lower than that of the BS mode. Whereas the vibration of either the apical or the BS mode raises the total energy, the vibrations of these two modes together produce an energy gain for the CDW state as compared with the vibration of each single mode. This energy gain indicates that the coupling of these two phonon modes with CDW is favorable; the coupling strength, measured by the energy gain, is found to be stronger than that with other periodicities, such as $6a$. The coupling strength also enhances with increasing hole concentrations.  These findings are in agreement with the RIXS results that strong phonon softening occurs about $q_{\|}=0.25$. One plausible scenario is that the CDW has effectively a period $4a$ and is short-ranged \cite{MesarosPNAS2016}; a shift in the wave vector of a large sample appears in the Fourier space, giving rise to the formation of discommensurate CDW. 

\section{Discussion and Conclusion} 

Although the data presented in the present work is only from a sample near the optimal doping, quantum critical scalings of charge fluctuations in temperature domain are clearly demonstrated. These results are consistent with the speculation on the existence of a QCP.  On the other hand, by suppressing superconductivity in high magnetic fields, linear scalings in resistivity have been observed in more extended regime around the putative QCP near doping level of 0.2 \cite{Cooper2009}. While the extended linear-resistivity region shows quantum scaling in temperature, it does not exhibit the quantum critical scaling in non-thermal parameter, doping. Thus, the correspondence between quantum criticality and T-linear resistivity may not be of necessity \cite{Ando2004,Phillips2005}. The apparent extended critical region that is overlapped with the putative QCP may just be a non-Fermi liquid phase \cite{doiron2003}. In this scenario, the CDW phase becomes a non-Fermi liquid phase as the doping level is tuned across the putative QCP. Clearly, to clarify these issues, a future work on samples (both LSCO and Nd-LSCO) with different dopings around the critical point and inside the region of long-range charge-ordered phase (Nd-LSCO sample) is required in order to unequivocally prove the existence of QCP related to charge order in LSCO.

To conclude, we observed short-range CDW in LSCO near the optimal doping over a wide temperature range; these orders compete locally with the superconducting order. Through combining high-resolution RIXS measurements, theoretical modeling and diagrammatic calculations, our results provide evidence for the phonon softening in superconducting cuprates induced by charge order. The observation of phonon softening indeed results from quantum fluctuations of the charge order.

\begin{acknowledgments}
We thank Wei-Sheng Lee and Alfred Baron for useful discussions. This work was supported in part by the Ministry of Science and Technology of Taiwan under Grant No. 109-2112-M-213 -010 -MY3 and 109-2923-M-213-001.  S.J. acknowledges support from the National Science Foundation under Grant DMR-1842056.
A.F.K. acknowledges support from the National Science Foundation under Grant No. DMR-1752713.
We also thank the support by KAKENHI Grant No. 15H02109 from JSPS and “Program for Promoting Researches on the Supercomputer Fugaku” (Basic Science for Emergence and Functionality in Quantum Matter) from MEXT.
\end{acknowledgments}


\bibliography{reference2}

\end{document}